\documentclass[letterpaper, 10 pt, conference]{ieeeconf}  

\IEEEoverridecommandlockouts                              

\usepackage{graphics} 
\usepackage{amsmath} 
\usepackage{amssymb}  
\usepackage{subfig}
\PassOptionsToPackage{hyphens}{url}\usepackage{hyperref}
\usepackage{dsfont}
\usepackage{algorithmicx}
\usepackage{algpseudocode}

\usepackage{todonotes}
\usepackage{soul}
\definecolor{smoothgreen}{rgb}{0.7,1,0.7}
\sethlcolor{smoothgreen}

\title{\LARGE \bf
Identifying Modes of Intent from Driver Behaviors \\ in Dynamic Environments
}

\author{Katherine Driggs-Campbell and Ruzena Bajcsy
\thanks{This material is based upon work supported by the National Science Foundation Award ECCS-1239323 and by the Office of Naval Research MURI Award ONR-N000141310341 and DURIP Award N000141310679.}
\thanks{
The study protocol was approved by the University of California Institutional Review Board for human protection and privacy, under Protocol ID 2013-07-5459. Each subject was first informed on the experimental procedure and written informed consent was obtained.}
\thanks{Katherine Driggs-Campbell and Ruzena Bajcsy are with the Department of Electrical Engineering and Computer Sciences, University of California at Berkeley, Berkeley, CA 94720 USA (e-mail: \{krdc,bajcsy\}@eecs.berkeley.edu).}%
}

\begin{document}

\maketitle
\thispagestyle{empty}
\pagestyle{empty}

\begin{abstract}
In light of growing attention of intelligent vehicle systems, we propose developing a driver model that uses a hybrid system formulation to capture the intent of the driver.
This model hopes to capture human driving behavior in a way that can be utilized by semi- and fully autonomous systems in heterogeneous environments.
We consider a discrete set of high level goals or intent modes, that is designed to encompass the decision making process of the human.
A driver model is derived using a dataset of lane changes collected in a realistic driving simulator, in which the driver actively labels data to give us insight into her intent.
By building the labeled dataset, we are able to utilize classification tools to build the driver model using features of based on her perception of the environment, and achieve high accuracy in identifying driver intent.
Multiple algorithms are presented and compared on the dataset, and a comparison of the varying behaviors between drivers is drawn.
Using this modeling methodology, we present a model that can be used to assess driver behaviors and to develop human-inspired safety metrics that can be utilized in intelligent vehicular systems.
\end{abstract}


\section{Introduction}\label{sec:intro}

In recent years, there has been a growing interest in intelligent vehicle systems, including advanced driver assistance, driver monitoring, and autonomous systems \cite{krdc2015,litman2015}.  
Many of these systems require a model of the driver, which has been shown to improve control algorithms and can be used to avoid safety heuristics \cite{shia2013}.  
These human centered systems are going to be increasingly important as more and more autonomy is introduced onto the roads \cite{little1997}.

In this paper, we consider a driver model that assesses driver intent to analyze what influences human decision making while driving.  
By developing this model, we hope to achieve two goals: (1) accurately capture a driver's intent in dynamic environments (i.e. the road properties and surrounding vehicles) and (2) design a system that is flexible and portable enough to be used in a variety of applications (e.g. driver feedback or autonomous decision making).
While humans are prone to distractions while driving \cite{regan2008driver}, we have many desirable qualities like flexibility, adaptability, and efficient high level decision making.  

Ideally, an autonomous system would be able to mimic these positive aspects of drivers while mitigating the drawbacks, and in doing so would improve social understanding and acceptance.
This is motivated by work in \cite{Yen2005}, as the authors demonstrated that sharing a mental model improves human-robot collaborations. 
In \cite{Becchio2012}, the authors showed that intent is directly tied to human motion, which others can detect and use to anticipate and understand intent; thus implying that understanding intent is ``deeply rooted'' to social interaction.

We propose a hybrid system inspire model that identifies the intent of the ego vehicle that can be used to assess driver behavior and/or assist in autonomous decision making in a heterogeneous environment (i.e. an environment with a mixture of human driven and autonomous vehicles).  
In order to ensure portability of the model to systems with mixed controllers, the detection relies on the sensor measurements of the environment surrounding the ego vehicle and not directly on the control inputs or driver state.  
By using a state of the art simulator, a dataset is collected with a real-time labeling system to gather sensor information from the environment and the human’s intent as she drives.  
Using this dataset, we are able to analyze the influences on decision making, assess a driver behaviors in different modes, and develop an algorithm that can accurately determine the intent of a driver based on the measurements of a dynamic environment to fit a hybrid model formulation.

This work is unique from other modeling methods for the following reasons: (1) a real-time, human labeled dataset of lane changes is collected, that is better able to capture the human decision making process and (2) the switching of modes are determined by the dynamics of the environment to better encompass the decision making process than previous methods, that identify modes by heuristics and time.

The paper is organized as follows.  In the subsequent section, we briefly discuss related works.  In Section \ref{sec:methods}, we present our modeling methodology.  Section \ref{sec:experiment} describes the experimental setup and Section \ref{sec:results} presents the results and analysis of the data and the corresponding models.  Finally, the results are discussed and summarized in Section \ref{sec:conclusion}. 

\subsection{Related Works}
\label{sec:related}

There are many works that consider predicting driver behavior by monitoring the driver \cite{krdc2015}, and have shown promising results in terms of human-in-the-loop and shared control for semi-autonomous frameworks \cite{shia2013}.
When it comes to driver modeling intent, many rely heavily on the driver state \cite{Doshi2011a} or on driver input \cite{Kuge2000}.
While effective, these methods rely on heuristics to determine when a lane change begins use windowing to select features that will train the classifier \cite{Doshi2011a}.
This ultimately assumes that these high level decisions are made as a function of time, and not by the dynamic state of the environment, which is difficult to predict due to the high variability over long time horizons \cite{shia2013}.

One of the desired outcomes of this work is to identify a model that does not depend on human input and can be used in human driven or autonomous systems. 
Ideally, the model could be used in human-inspired driving applications.
Driving styles in terms of discrete control actions were mimicked in \cite{abbeel2004}, using inverse reinforcement learning.

By requiring that the human driver explicitly label the current mode being driven, we can use supervised classification approaches to generate system identification parameters for these modes of intent. 
The resulting model based on observations of the surround vehicles that can be detected using current sensor technology, with features similar to the cues that humans perceive while making driving decisions.  
As previously mentioned, this allows the algorithm to be used as a decision system for an autonomous system or an assistance system for a human driven or semi-autonomous vehicle that can effectively function in a mixed environment.

\section{Methods}\label{sec:methods}

In this work, we model each mode of intent as a discrete state, which has a different controller (or control objective) associated with that mode.  
As the driver navigates through the environment, she transitions between the modes of intent, switching controllers in each mode. 
We assume that the switching of modes is only influenced by the state of the surrounding environment (i.e. the driver will not spontaneously change lanes when there are no other cars nearby).  

\subsection{Hybrid System Approach}

A hybrid system is a representation of a dynamical system that has continuous dynamics that depend upon a discrete state or mode of operation.
Suppose we are given the vehicle dynamics for the ego vehicle, which can be approximated using a bicycle tire model extended with roll dynamics, as found in \cite{rajamani2005}.  
We compactly represent these dynamics in the following discretized form:
\begin{equation}
x_{k+1} = f(x_k,u_k), \quad \forall k \in \mathbb{N} 
\end{equation}
where $x_k \in \mathbb{R}^n$ is the state of the vehicle and $u_k \in U$ is the input from the input space at time $k$.
The following states are those of interest:
\begin{equation}
x_{k} = \begin{bmatrix}
p_x & p_y & v_x & v_y & \theta 
\end{bmatrix}^\top
\label{eq:states}
\end{equation}
where $p_x$ and $p_y$ are the positions with corresponding velocities $v_x$ and $v_y$, and $\theta$ is the heading angle of the vehicle.  
We can decompose the dynamics into the following form:
\begin{equation}
x_{k+1} = f(x_k)+g_q(x_k,u_k), \quad \forall k \in \mathbb{N} 
\label{eq:veh_dyn}
\end{equation}
where all variables are as before and $g_q(x_k,u_k)$ is the control law that defines the input to the vehicle that changes with the associated mode $q \in Q$.
If human driven, this assumes that the behaviors or inputs of the driver will change dynamically based on intent.
If semi- or fully autonomous, this algorithm might update the high level objective, constraints, or cost function for a control algorithm like model predictive control or would identify a new target set for control schemes using motion planning.

We suppose that the vehicle will be acting in a dynamic environment that will determine the mode of the ego car.
In our scenario, the set of modes $Q$ describe the various ``types" of driving that occur during normal driving.  
These might include lane changing, emergency avoidance maneuvers, lane keeping, etc.

However, as previously stated, we assume that the mode is determined not by the dynamics themselves, but by the observable, relative dynamics of the surrounding vehicles, which can be obtained via sensor measurements that are updated at each time step.  Suppose the sensors map the true dynamics of the environment to a noisy estimate through some function $h$:
\begin{equation}
\hat{y}_k^i = h(x_k^i) \quad \forall i = \{ 1,\dots,m \}
\label{eq:env}
\end{equation}
where $x_k^i$ is the true state (as in Eq. \ref{eq:states}) and $\hat{y}_k^i$ is the measured representation of vehicle $i$, supposing that we can detect and measure $m$ vehicles within a predefined radius of the ego vehicle.  
Here, we suppose $\hat{y}_k^i$ for $i = 1,\dots,m$ can be passed to a detection algorithm, $\mathcal{A}$, to calculate $\sigma_q \in \Sigma$, which is the set of discrete inputs that detects and signals changes in the environment that instigate a change of modes. Formally,
\begin{equation}
\sigma_q = \mathcal{A}(x_k,\hat{y}_k^i,\dots,\hat{y}_k^m)
\label{eq:modes}
\end{equation} 
where all variables are as previously described.
Note that $\mathcal{A}$ will compute and select features from the vehicle states, and will exclude control inputs to maintain flexibility.

\subsection{Modes of Intent}
In this paper, we focus on identifying the transitions between driver intent modes that a driver encounters while driving.
We pre-define the modes of behavior, collect the driver data to build a dataset, and learn the hybrid system model inspired by human driving. 

\begin{figure}[!ht]
\begin{center}
\includegraphics[height=2.25cm]{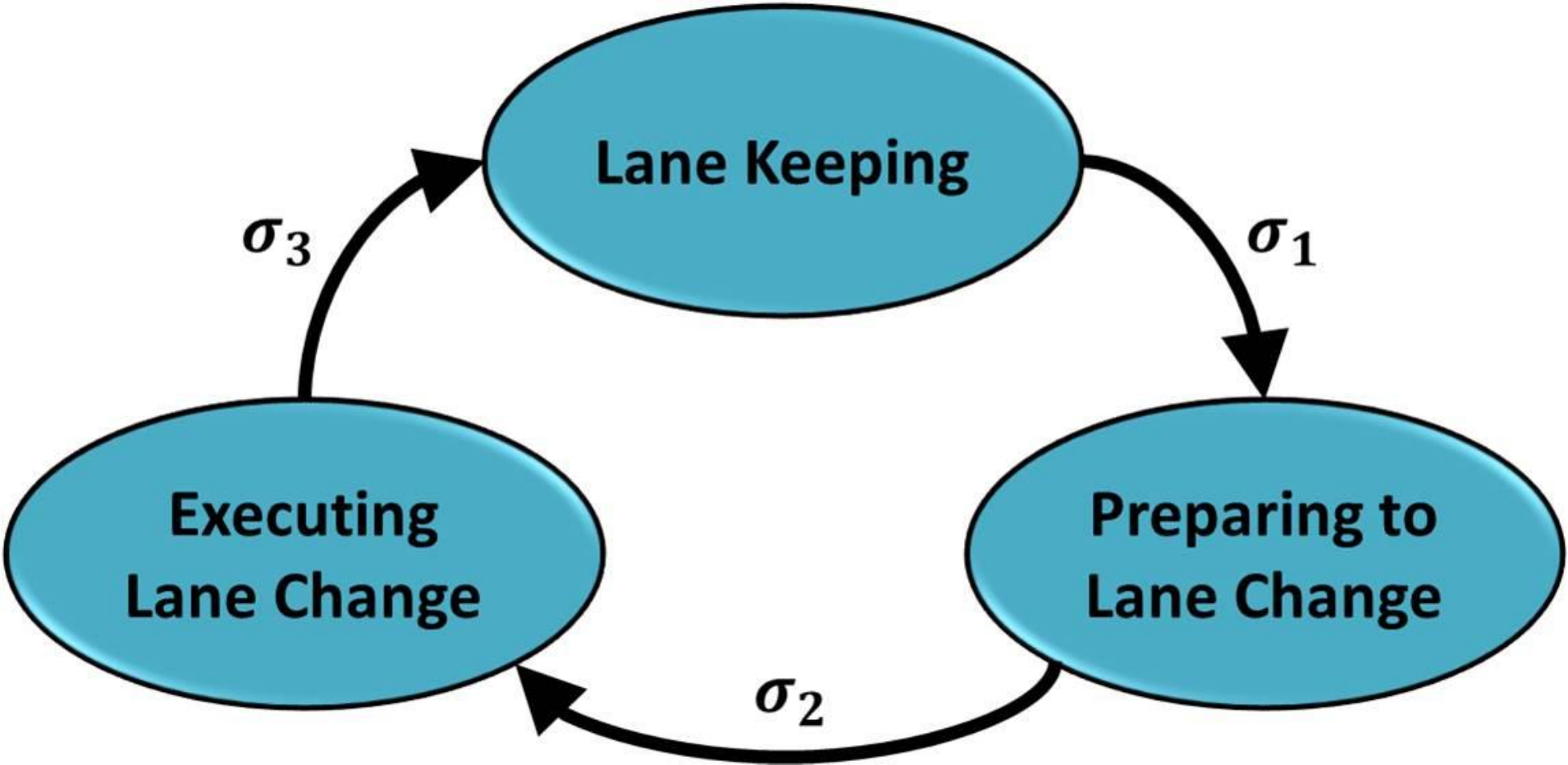}
\caption{\small Illustration of discrete modes in our hybrid model of driver intent, where we model the transitions as discrete inputs, $\sigma_q$.}
\label{fig:modes}
\end{center}
\end{figure}

For simplicity, we examine the scenario of driving in a two-lane, one way road, in a non-urban setting.  
We define three modes: \emph{lane keeping, preparing to lane change,} and \emph{lane changing}.
The model is visualized in Figure \ref{fig:modes}.
While \emph{lane keeping} and \emph{lane changing} are self-explanatory, \emph{preparing to lane change} can be thought of as the mode when the driver begins planning to change lanes and is waiting for the proper moment.
This can be thought of as the moment when the driver turns on their turning signal or blinker.
By identifying this mode, we inherently have a predictive nature in the system, by determining the lane change prior to the manuever actually occurring.
But instead of putting a distinct time stamp on the prediction horizon, we suppose that this relies on new instances on the environment.
In Section \ref{sec:results}, we show the empirical results for the prediction time horizon.

We introduce this in-between mode as a precursor to executing the lane change for two reasons.
First, detecting lane changes from a dataset has traditionally been done by determining when a lane change occurs by some heuristic (e.g. when the heading angle passes a particular threshold or when the vehicle exits the lane).  
These models look at the data leading up to this point in order to predict that a lane change will occur in the next few seconds \cite{Doshi2011a,Kuge2000}.
This, however, does not capture the decision making process of the human, or capture the idea that these decisions occur as a function of the environment, not just time.

Second, we introduce this in-between state in hopes to capture the variation that humans exhibit as their intent change.
As humans drive, we are constantly assessing the environment, and building a mental model of what is happening around us by determining the intent of other vehicles.
While we often rely on turning indicators or blinkers to convey our intent to surrounding vehicles, humans can often determine whether or not a vehicle wants to change lanes without the signal.
If we assume that vehicles are to be integrated into society, we can assume that they will need to be predictable and convey their intent to human drivers beyond just using the indicator.

This is helpful in building our system, because (1) it mimics the decision making process more accurately by introducing an intermediary step between the two obvious actions and (2) it allows the model to varying the amount of prediction for when the driver would change lanes, by relying on the state of the environment dynamics, not a predetermined time horizon.

To learn these transitions or discrete inputs (Eq. \ref{eq:modes}), we build a dataset that consists of the environment, as represented in Equation \ref{eq:env}, with corresponding mode labels.
Details about the experiments and data collection are presented in Section \ref{sec:experiment}.
Since we have collected a set of labeled data, we may use supervised classification techniques to learn and analyze the transitions between modes.

\subsection{Features from the Dataset}

As described, we were interested in analyzing the effects of the dynamic environment on driver's intent and decisions.
Therefore, we utilized the observable and measurable dynamics of the surrounding vehicles given current sensing technology as the starting point of the features. 
This means that from a dataset of $x$ and $\hat{y}^i$ we generate features denoted as elements from the feature space as $z \in \mathcal{F}$, with associated label $q \in Q$ which match the predefined modes. 
The distances to the surrounding vehicles are the primary raw features, as relying on heading angle and inputs would eliminate the portability of the algorithm.

To make the features robust to changes in position, velocity, and control input, we considered relative positions and velocities, heading relative to the road, and as well as two time metrics: time-to-collision (TTC) and time headway (THW), which are commonly  used as a metric for threat  and driver perception \cite{abbink2008}, as well as the relative TTC and THW, denoted rTTC and rTHW.
These are defined as follows:
\begin{equation}
\begin{aligned}
\text{TTC}_i = \frac{d_i}{v_e} \;, \qquad 
\text{THW}_i = \frac{v_e}{d_i}
\label{eq:ttc}
\end{aligned}
\end{equation}
\begin{equation}
\begin{aligned}
\text{rTTC}_i = \frac{d_i}{v_i} \;, \qquad 
\text{rTHW}_i = \frac{v_i}{d_i}
\label{eq:rttc}
\end{aligned}
\end{equation}
where $v_e$ is the speed of the ego vehicle and $d_i$ and $v_i$ are the relative position and velocity, respectively, to vehicle $i$.

We define a detection region of $50m$ for which features of detected vehicles are included.  
For consistency, the feature vector is ordered such that we consider a small grid around the ego vehicle.
The first position in front of the ego vehicle and in the same lane. 
The second position is the in front of the mid-way point of the ego vehicle and in the opposite lane.
The third and final position is in the opposite lane as the ego vehicle, behind the mid-way point. 
This grid is visualized in Fig. \ref{fig:order}, and is mirrored when the ego vehicle is in the left lane.
If no vehicle is in range in one of the three defined position, padding is inserted to maintain 
ordering\footnote{Let it be noted that a variety of feature sets were tested in developing this model, including one that developed different classifiers depending on the number of surrounding vehicles that were in the detection region.  It was found that the effectiveness of the model did not change if multiple models were used or if the features were concatenated as is presented here.

We'll also take this time to note that the labels or modes were varied and tested, by using the three modes presented here but labeled for the left and right lane, creating a total of six modes.  While this showed promising results, to keep the model simple, we limited the number of modes.  
Additionally, preliminary studies show promising results that this can easily be extended to roads with more lanes and of varying curvature.  However, as the scenarios expand more sufficient data must be collect as well as more modes.
}.
Using the normalized feature vector and corresponding labels, we use classification techniques to identify the driver model.

\subsection{Identifying Transitions}
As was mentioned, we translate this problem to a detection or classification problem, using driver data.
Assuming we have a complete, labeled data set of human behaviors, there are many supervised methods that will be able to accept the sensor measurements as features, and return the expected mode.
We present results using the following techniques: 
Support Vector Machines (SVM), Random Forests (RF), and Logistic Regression (LR).

SVM and RF were tested due to their efficiency, their robustness to overfitting, and the fact that are a logical choice when finding regions associated with classes or when identifying separating hyperplanes.  
LR was tested as it is the most popular of soft classification algorithms, meaning it assigns a score or a probability that a sample belongs to a class given the features. 
This should be able to capture the ``gray" area that is present in human decision making by identifying the likelihood or mixture of the modes in a given sample.
For further information, we guide the reader to \cite{hastie2009}.

\section{Experimental Setup}\label{sec:experiment}

In this section, we describe our experimental setup for collecting the dataset and the features used to discriminate modes. 
When studying human-in-the-loop systems, one of the challenges is collecting realistic data in a safe manner.  
To address this, we have developed an experimental setup for studying human-in-the-loop systems in vehicles, particularly in driving applications.  
The resulting testbed was designed to be a flexible, realistic platform that allows us to both observe the driver with monitoring devices, but also control and measure the environment \cite{driggscampbell2014}.

\begin{figure}[!t]
\centering 
\subfloat[Ordering]{\includegraphics[height=3cm]{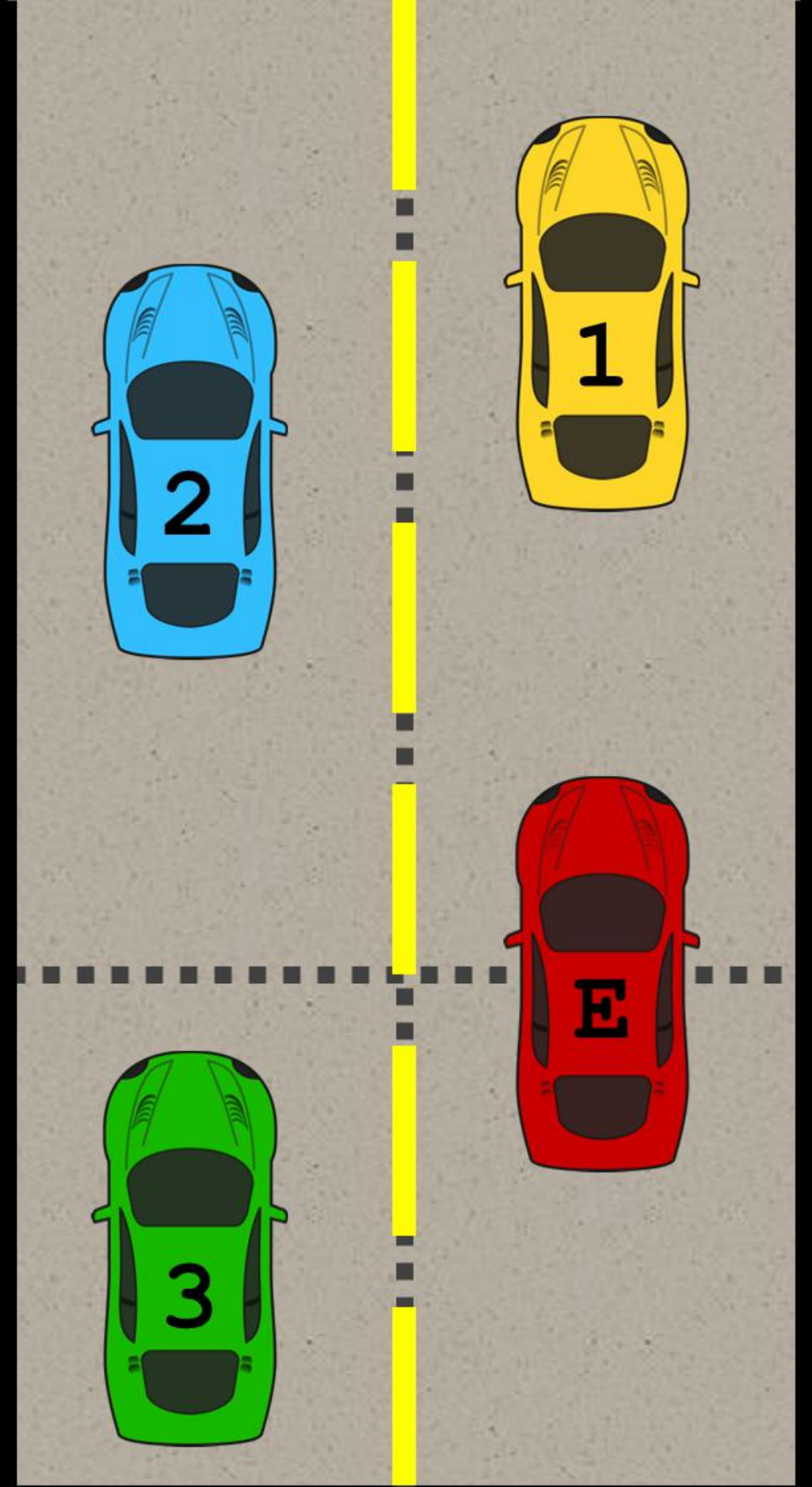}
\label{fig:order}}
\subfloat[Test subject driving the simulator]{\includegraphics[height=3cm]{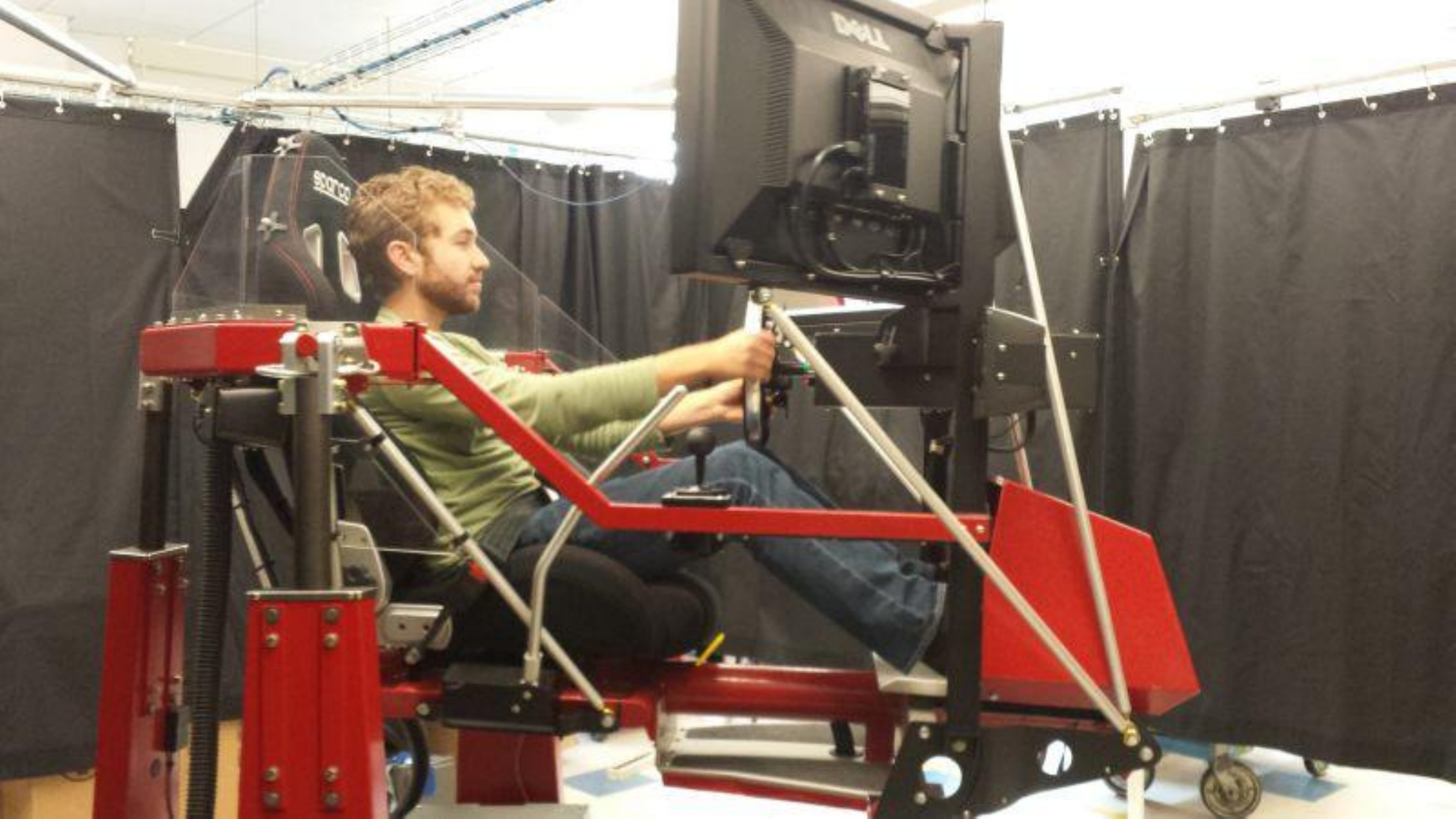}
\label{fig:sim_exp}}
\caption{\small (a) Illustration of the ordering of feature vector depending on location of vehicle relative to the ego vehicle (denoted E).  (b) Picture of the driver in the motion platform simulator. 
}
\label{fig:simulator}
\end{figure}

Driver data was collected using a Force Dynamics CR401, a 4-axis motion platform simulator, which recreates the forces experienced while driving \cite{force_dynamics}.  
This system has been integrated with PreScan software, which provides vehicle dynamics and customizable driving environments allowing us to recreate various driving environments needed for data collection \cite{prescan}.  
Using this human-in-the-loop test bed, we are able to reliably and realistically obtain driver data that can illustrate the utility of our models and provide useful motion feedback to the drivers.  
The motion simulator and visualization seen by the driver is shown in Fig. \ref{fig:sim_exp}.

To set up the experiment, multiple scenarios were created in which the driver traverses a straight two lane road attempting to maintain a speed between $15$ and $20$ m/s.
Scenarios were generated by creating combinations of the simulation parameters to collected a complete dataset.  
The following parameters were varied:
(1) the initial speed and lane location of ego vehicle;
(2) the number and location of surrounding vehicles, varied from one to three; and 
(3) the initial and final speed of each surrounding vehicle.

For example, in some scenarios, the lead vehicle would slow down, forcing the driver to change lanes only if there was room in the next lane. 
Thus, the key here is finding the configurations of the environment states that cross the boundary or safety margin of the human and allows us to identify their likely action between staying in the lane (i.e. braking) or changing lanes to maintain her desired speed.

While driving, the human driver actively labels the data into classes or modes for the training and testing sets during the experiment. 
The driver presses a button on the steering wheel to signal that she was preparing to lane change and then presses a paddle on the steering wheel to indicate that she was executing the lane change. 
This allowed separate the data into the modes as previously mentioned.

Five drivers were recruited for this experiment.  
The driver's were asked to drive for an hour on two separate occasions to collect the data. 
An optional practice session was offered, to get the driver accustomed to the simulator and the data collection method. 
The dataset ultimately resulted in about 200 lane changes, per driver, which was divided into two datasets for training and testing.
We note that some scenarios required did not require a lane change (e.g. the relative speed of the lead vehicle was initialized such that the driver never felt the need to overtake them), while other scenarios which heavy traffic caused multiple lane changes, but varied depending on the driver's behaviors in the simulation. 
The vehicle dynamics, driver mode, and environment is sampled at the synchronized rate of 60 Hz.

\section{Results}\label{sec:results}

In this section, we examine the influence of the features on driver behavior and describe the models developed from the aforementioned algorithms.

\subsection{Analyzing Driver Behaviors}

From this dataset that describes driver intent, we are able to visualize the variability in human driving and gain insight to the connection between a driver's perception of her surrounds to her discrete state of intent.
This data can also verify the utility of introducing the \emph{preparing to lane change} mode; the advantage of the human active labeling method over the traditional heuristic labels and windowing; and compare the differences in behaviors between drivers.

\begin{figure}[!t]
\begin{center}
\includegraphics[height=5cm]{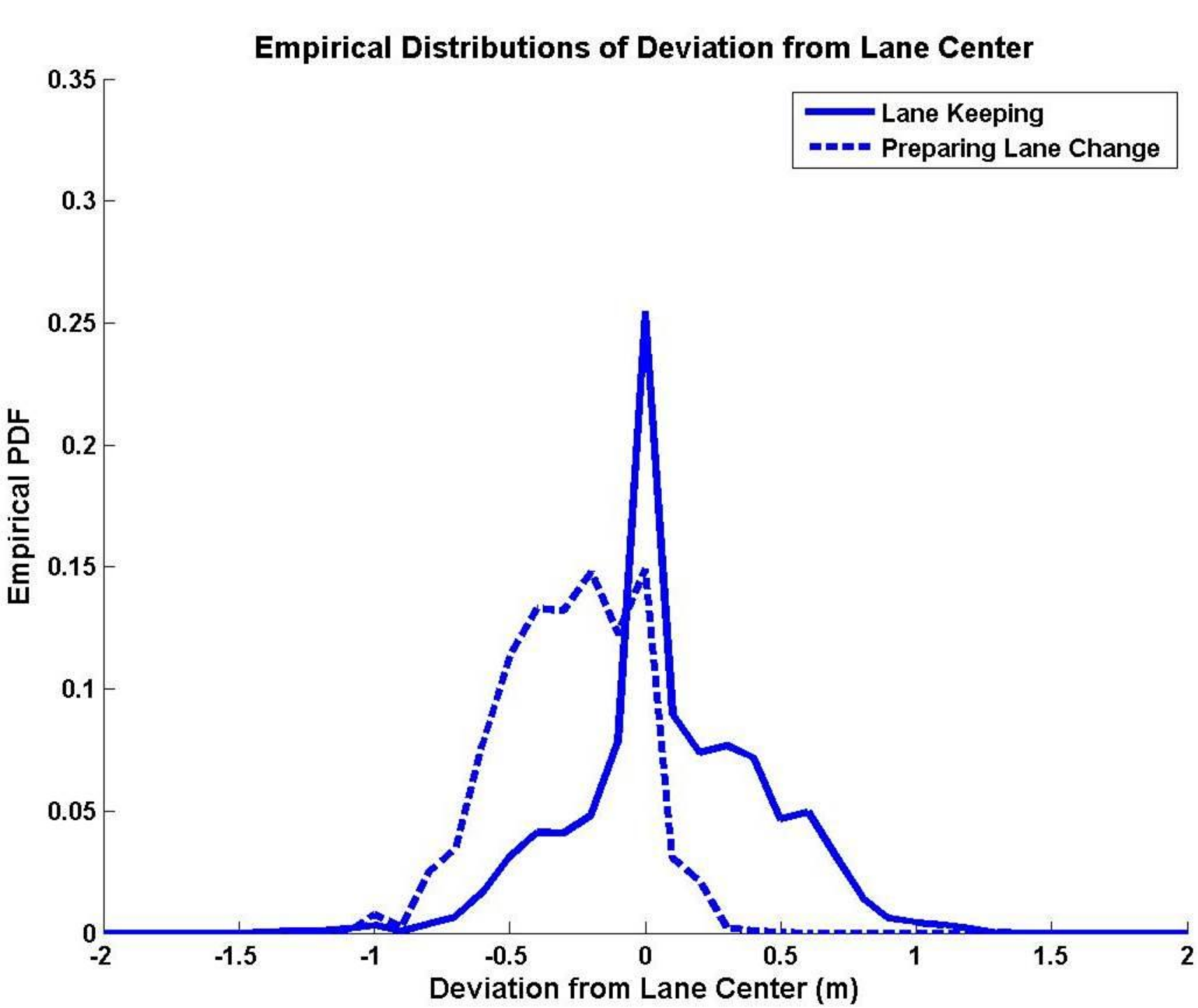}
\caption{\small Empirical probability density functions that describe the different driving behaviors exhibited in the similar, yet distinct, lane keeping and preparing to lane change modes, for a particular driver. This data consists of data from the right lane, but we noted similar, but reflected behaviors, in the other lane.}
\label{fig:prep}
\end{center}
\end{figure}

To justify separating the lane keeping modes into two modes, we consider the distribution of lateral deviations from the center of a lane, which is assumed to be the driver's internal goal in this mode. 
We observe that drivers tend to distribute themselves closer to the center lane and are more likely to edge away from traffic when simply lane keeping, and move toward next lane when preparing to lane change.
The empirical distributions of these behaviors are shown in Figure 
\ref{fig:prep}
\footnote{The subsets of the data were compared using the two-sample Kolmogorov-Smirnov test to see if they came from the same distribution.  The hypothesis was rejected with $p\ll 0.01$.}. 

\begin{figure}[!t]
\begin{center}
\includegraphics[height=5cm]{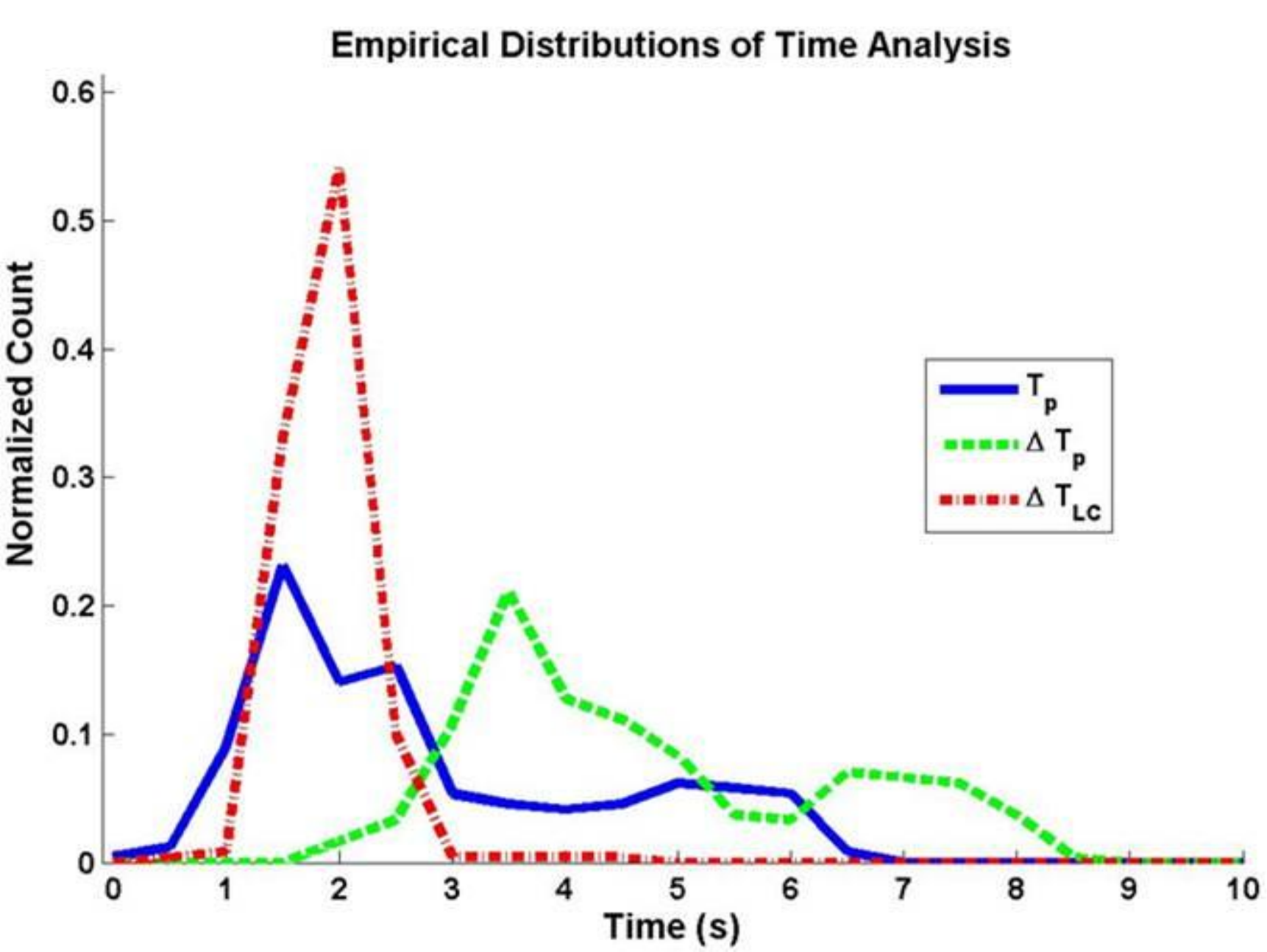}
\caption{\small Empirical probability distributions showing our time analysis metrics, where $T_P$ is the time spend preparing to lane change and $\Delta T$ is the change in time from when the driver entered the specified mode compared to the baseline. The mean and standard deviations for all drivers are shown in Table \ref{tab:results}.}
\label{fig:timing}
\end{center}
\end{figure}

To analyze the behavior in modes and across drivers, we also consider the time spent in the \emph{prepare} mode, $T_P$, as well as the time difference, $\Delta T$, between when the instance that human identified a mode transition and the instance that the lane change occurred according to a baseline heuristic, which was determined as the time the bounding box of the vehicle exited the lane.
This evaluation is presented in Figure \ref{fig:timing} for a single driver.

By visualizing these time metrics, we observe that the time between modes varies a great deal, implying that the decision making process does not solely depend on time, but on the dynamics of the environment.
This also empirically shows the predictive capability of this model, showing that we are able to predict the intent of the vehicle, prior to the maneuver actually occurring. 
Similarly, the TTC and THW metrics can also be analyzed to identify the typical thresholds for the mode transitions as labeled by a driver.  
The averages for the timing metrics and these features for all subjects are presented in Table \ref{tab:results}. 

\begin{table}[t!]
\centering
\caption{\small Results from the timing analysis and the TTC and THW metrics are presented, showing the mean and standard deviation of the means for all subjects.  Subscript $P$ denotes result for \emph{prepare} mode and $LC$ denotes \emph{lane change} mode.}
\begin{tabular}{|l|c|c|}
\hline
{\bf Metric} & {\bf Mean (s)} & {\bf St. Dev. (s)} \\
\hline \hline
$T_{P}$	   & 3.05 & 0.29  \\
\hline
$\Delta T_{P}$   &  2.66 & 1.17 \\ 
\hline
$\Delta T_{LC}$  &  1.52 & 0.28 \\ 
\hline \hline
$TTC_{P}$  & 1.34 & 0.17 \\
\hline
$TTC_{LC}$ & 1.20 & 0.14 \\ 
\hline \hline
$THW_{P}$  & 0.80 & 0.11 \\ 
\hline
$THW_{LC}$ & 0.96 & 0.12 \\ 
\hline
\end{tabular}
\label{tab:results}
\end{table}

\subsection{Model Performance}

It is worth noting that we cannot expect a perfect model, as there is uncertainty near the boundaries of the human's decision making process and noise in the actual labeling of data. 
Thus, errors will likely occur at the boundaries, but we expect the model to identify the optimal separating hyperplane to balance the noisy data.

\begin{figure}[!t]
\centering
\includegraphics[height=5cm]{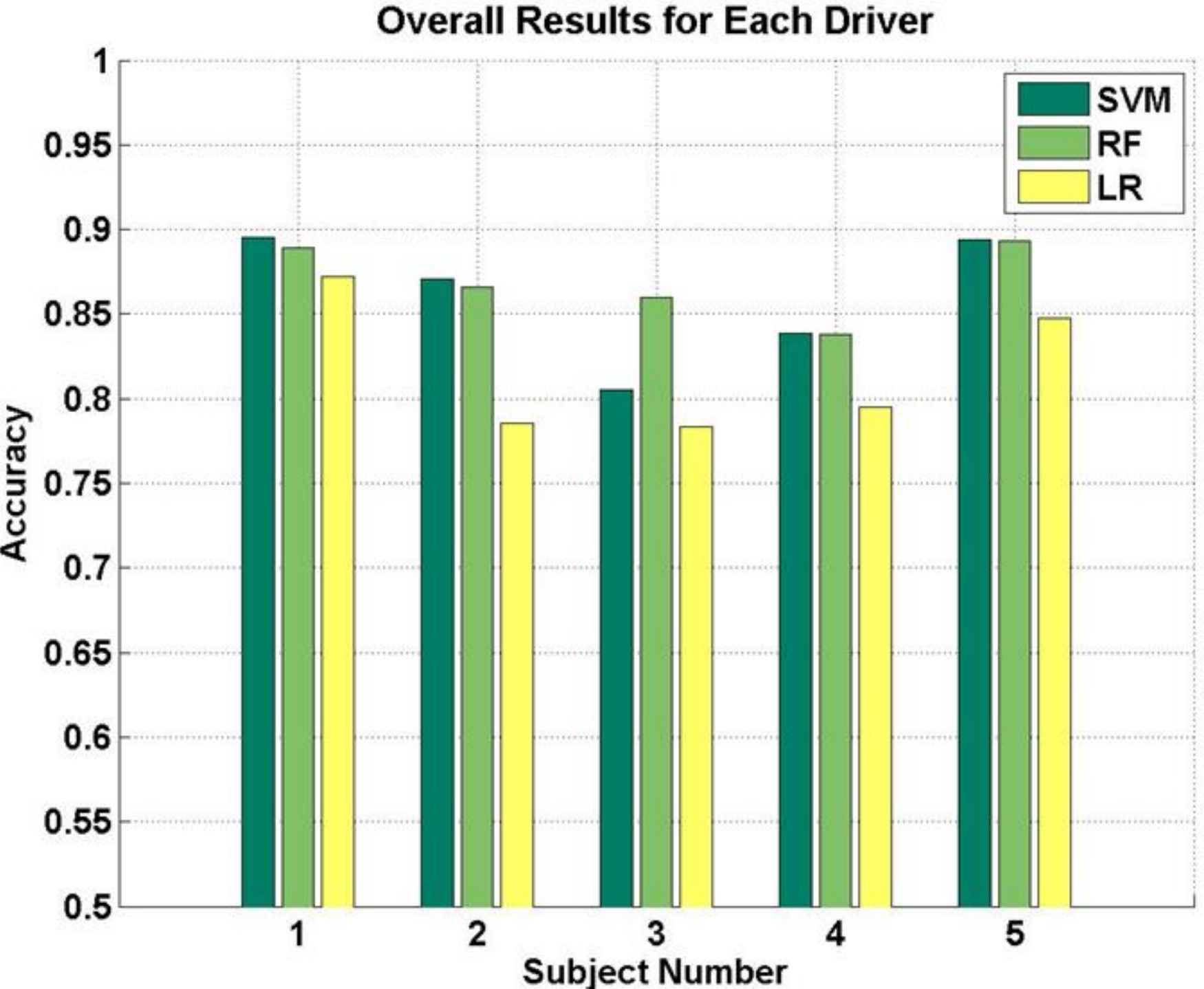}
\caption{\small Plot showing the accuracy for each of the five subjects.}
\label{fig:acc}
\end{figure}

By using the obtained parameters from cross validation and applying the generated classifiers to testing data to validate the model, we obtain the performance results presented Figure \ref{fig:acc} on the entire dataset.
Sample results for a single driver on the test dataset are shown in Table \ref{table:accuracy}. 
The accuracy is presented as the overall accuracy, and by scenarios that are defined by the number of vehicles in the immediate vicinity of the ego vehicle.

\begin{table}[!t]
\caption{\small This table presents the overall model accuracy (in \%) and the accuracy in different scenarios, denoted Scen. $i$, where $i$ indicates the number of vehicles in a 50m radius of the ego vehicle, for a particular driver only on the test dataset.}
\begin{center}
\begin{tabular}{|l||c|c|c|c|} 
\hline
\textbf{Method} & \textbf{Overall} & \textbf{Scen. 1} & \textbf{Scen. 2} & \textbf{Scen. 3} \\
\hline \hline
\textbf{SVM} & 89.5 & 88.5 & 	 85.9 & 	 92.1  \\
\hline
\textbf{RF}  & 88.9 & 86.6 & 	 85.9 & 	 91.7  \\ 
\hline
\textbf{LR} & 87.2 &	86.1 & 	 85.4 & 	 88.7  \\ 
\hline
\end{tabular}
\end{center}
\label{table:accuracy}
\end{table}

\begin{figure}[!t]
\includegraphics[width=.48\textwidth]{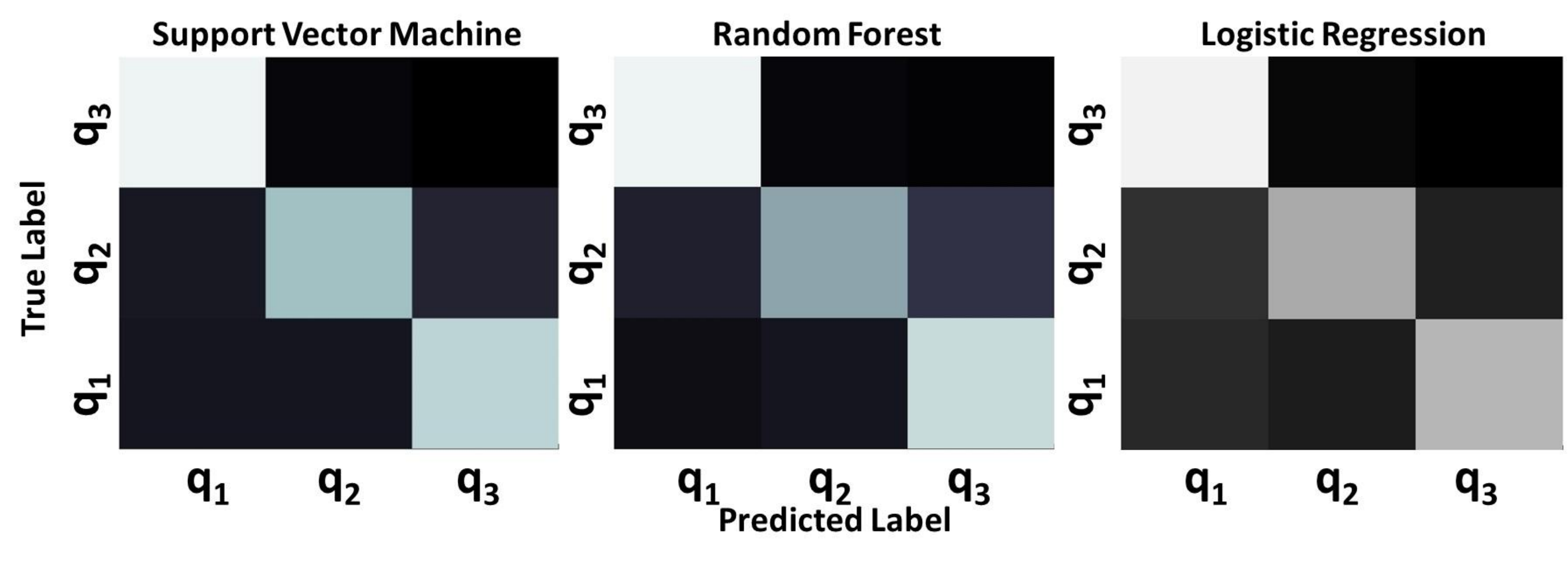}
\caption{\small Confusion matrices showing normalized accuracy results, on the test dataset for a given driver.  Entry $i,j$ shows the ratio of samples that are from mode $q_i$, but predicted by the algorithm to be $q_j$.  White indicates 100\% whereas black indicates 0\% accuracy.}
\label{fig:ROC}
\end{figure}

We can see that the generated classifiers determines the driver mode with relatively high accuracy, despite the uncertainty that is inherent in human actions.
To visualize the classification accuracy without biases in the number of instances, the confusion matrices showing the normalized results are shown below in Figure \ref{fig:ROC}.

As shown, this methodology shows promising results for identifying a hybrid model of an individual driver's decision making process.
The following observations were made:
\subsubsection{SVM}
We note that it is generally conservative, transitioning from the lane keeping modes later than the other methods, but is still able to predict lane changes prior before the maneuver occurs.  
This method also provides smoother transitions than the other methods.
This is to be expected given the method SVM balances the separating hyperplanes with the noisy labels.  

\subsubsection{RF}
This classifier can accurately distinguish the behavior modes, but is not always smooth about the borders, which is intuitive due to fine parsing the algorithm exhibits when identifying the separation boundaries.
While in theory this method is robust to overfitting, we noted an increase in performance between the test dataset and the complete dataset.
This implies that this method is more reliable under distinct operating conditions and does not generalize as well as the previous method.

\subsubsection{LR}
\begin{figure}[!ht]
\begin{center}
\includegraphics[height=4.25cm]{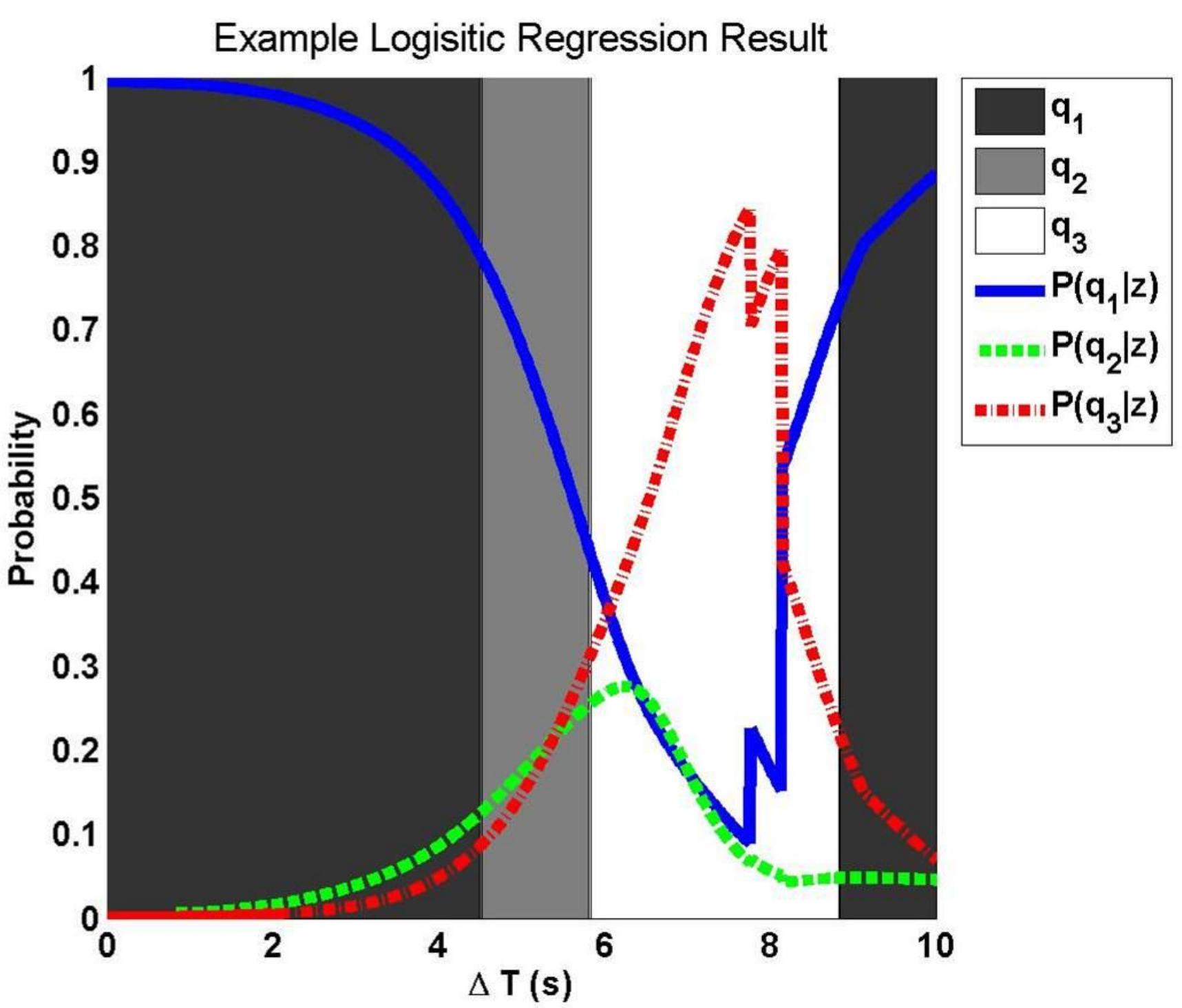}
\caption{\small Example illustration of how the logistic regression changes over the execution of a lane change, where the shaded regions show the actual mode, and the lines show the probability of a mode conditioned on the input features at that timestep.}
\label{fig:LR}
\end{center}
\end{figure}

The accuracy of this method was determined by using the most likely mode as the classification.
While it can be seen that this method produced relatively low accuracy, it is interesting as it shows the modal mixtures near the transitions between modes as shown in Figure \ref{fig:LR}. 

\subsubsection{Overall Performance}
We note that the accuracy of the classification varies a great deal between subjects, which is likely due to the inaccuracies of human labeling and the fuzzy nature of human decisions.
When the data is combined in an attempt to generate a global model, extremely poor performance was found.
If this method were to be used to in practice, an expert driver would be selected (i.e. one that generates a reliable model).
This expert could be used as a standard to compare other drivers' behaviors to or to generate a high level control algorithm for use in semi- or fully autonomous vehicles.

Additionally, no windowing was used, although it is common practice in similar studies (see Section \ref{sec:related}).
While this might improve the accuracy, as reasonable results are found without windowing it was decided that using minimal features was desirable, to maintain flexibility and eliminate reliance on the trajectories generated on the simulator.

\section{Discussion}\label{sec:conclusion}

We have compiled a dataset of human labeled lane changes and developed a hybrid mental model to mimic a driver's decision making process, by learning a detection module to identify the separating hyperplane between driver modes. 
This model is unique as it provides a robust classification of human decisions assuming that the transitions are not determined by time, but by the dynamic environment, and uses a driver labeled dataset.
The models have been shown to be highly robust to variations in the environment and exhibited high accuracy.
Future work consists of expanding the dataset to consider more scenarios and more complex roadways, that will allow for the surrounding vehicles to display more uncertain and variable behaviors.  
We would also like verify the portability to other vehicles and work towards implementing this work as a high-level controller for autonomous systems in heterogeneous environments.

\bibliographystyle{abbrv}
\bibliography{krdcBibFile}

\end{document}